\documentclass[useAMS,usenatbib]{mn2e}
\usepackage{graphicx}
\usepackage{amsmath}

\title{A CSO Search for $l$-C$_3$H$^+$: Detection in the Orion Bar PDR}
\author[B. A. McGuire et al.]
{Brett A. McGuire$^1$, 
P. Brandon Carroll$^1$, 
James L. Sanders III$^2$,
\newauthor Susanna L. Widicus Weaver$^2$, 
Geoffrey A. Blake$^{1,3}$, 
and Anthony J. Remijan$^4$ \\
$^1$Division of Chemistry and Chemical Engineering, California Institute of Technology, Pasadena, CA 91125, USA\\
$^2$Department of Chemistry, Emory University, Atlanta, GA 30322, USA\\
$^3$Division of Geological and Planetary Science, California Institute of Technology, Pasadena, CA 91125, USA\\
$^4$National Radio Astronomy Observatory, Charlottesville, VA 22903, USA}

\begin{document}

\pagerange{\pageref{firstpage}--\pageref{lastpage}} \pubyear{2014}

\maketitle

\label{firstpage}

\begin{abstract}
The results of a Caltech Submillimeter Observatory (CSO) search for $l$-C$_3$H$^+$, first detected by \citet{Pety2012} in observations toward the Horsehead photodissociation region (PDR), are presented.  A total of 39 sources were observed in the 1 mm window. Evidence of emission from $l$-C$_3$H$^+$ is found in only a single source - the Orion Bar PDR region, which shows a rotational temperature of 178(13) K and a column density of $7(2) \times 10^{11}$~cm$^{-2}$. In the remaining sources, upper limits of $\sim$10$^{11} - 10^{13}$~cm$^{-2}$ are found.  These results are discussed in the context of guiding future observational searches for this species.
\end{abstract}


\section{Introduction}
\label{intro}

\citet{Pety2012} have reported the detection of eight transitions of a closed-shell, linear molecule in observations toward the Horsehead photodissociation region (PDR).  They performed a spectroscopic analysis and fit to these transitions frequencies and, based on comparison with the theoretical work (see \citealt{Ikuta1997} and refs. therein), attribute these transitions to the $l$-C$_3$H$^+$ cation.  Later, \citet{McGuire2013a} identified the $J = 1 - 0$ and $J = 2 - 1$ transitions predicted by \citet{Pety2012} in absorption toward the Sgr B2(N) molecular cloud, as well as tenuous evidence toward Sgr B2(OH) and TMC-1.  The attribution of these signals to the $l$-C$_3$H$^+$ cation was later disputed by \citet{Huang2013} and \citet{Fortenberry2013}, with the latter suggesting the anion, C$_3$H$^-$, as a more probable carrier based on high-level theoretical work.  \citet{McGuire2014} found that the observational evidence at that time supported the assignment of this carrier to $l$-C$_3$H$^+$.  Recently, \citet{Brunken2014} reported the first laboratory measurements of $l$-C$_3$H$^+$ and confirmed the astronomical assignment.

While the question of identity has now been resolved, questions remain surrounding the formation conditions and chemical implications of $l$-C$_3$H$^+$.  Because $l$-C$_3$H$^+$ has been definitively detected in only two environments - the Horsehead PDR and Sgr B2(N) - efforts to explore these questions are hampered by a lack of information.  In an attempt to address this deficiency, we have conducted a wide search of PDRs and complex molecular sources in search of $l$-C$_3$H$^+$.  Here, we present the results of a brief, targeted campaign of 14 astronomical sources with the Caltech Submillimeter Observatory (CSO) covering the $J = 10 - 9$ and $J = 12 - 11$ transitions of $l$-C$_3$H$^+$.  We  also examine the $J = 10 - 9$ transition in broadband unbiased line surveys of a further 25 sources.    The observational details are given in \S\ref{obs}, resulting spectra are presented  and data reduction strategies are outlined in \S\ref{results}, and a discussion follows in \S\ref{discussion}.

\section{Observations and Data Reduction}
\label{obs}

Observations as part of the targeted campaign to detect $l$-C$_3$H$^+$ were conducted over the course of 4 nights in May 2013, 2 nights in October 2013, 2 nights in November 2013, and 2 nights in January 2014 as part of early remote observing trials using the CSO.  The dataset of 25 unbiased molecular line surveys was obtained with the CSO between September 2007 and June 2013 in the frequency region of the $J = 10 - 9$ transition.  

\subsection{Targeted Campaign}
\label{bamcso}

The CSO 230/460 GHz double side band (DSB) heterodyne sidecab receiver, operating in its 210 - 290 GHz mode, was used in moderately good weather ($\tau \sim 0.07-0.12$) resulting in typical system temperatures of $T_{sys} \sim 250$ K.  The backend consisted of two Fast Fourier Transform Spectrometers (FFTS): FFTS1 provided 1 GHz of DSB spectra at 122 kHz resolution while FFTS2 provided two, 2 GHz DSB spectral windows at 269 kHZ resolution.   For the Orion Bar observations, the receiver was additionally used in its 170 - 210 GHz mode to observe the $J = 9-8$ transition at 202 GHz.  The $J = 8 - 7$ transition at 180 GHz was not observed due to interference from the nearby water line.

Target sources and parameters are given in Table \ref{bamtargetsources}.  For observations of sources with known extended structure, position switching observations were used.  For more compact sources, a chopping secondary mirror, with a throw of 2$\arcmin$ was used - this resulted in lower overhead times than position switching observations.  Details are given in Table \ref{obssummary}. The raw data were intensity calibrated using the standard chopper wheel calibration method, which placed the intensities on the atmosphere-corrected temperature scale, $T_a^{\ast}$.  All intensities were then set to the main beam temperature scale, $T_{mb}$, where $T_{mb}=T_a^{\ast}/\eta_{mb}$; the main beam efficiency was taken as $\eta_{mb}$ = 70\% for these observations. Pointing was performed every $\sim$2 hours, usually on a planetary source, with pointing corrections converging to within $\sim$1\arcsec.

\begin{table*}
\centering
\begin{minipage}{110mm}
\caption{Sources, coordinates, $V_{LSR}$, and source type for the targeted search.}
\begin{tabular}{l r r c l}
\hline
Source					&	$\alpha$(J2000)			&	$\delta$(J2000)					&	$V_{LSR}$ (km s$^{-1}$)		&	Notes						\\
\hline
Sgr B2(OH)				&	17:47:20.8				&	-28:23:32						&	64						&	Galactic Center, Hot Core			\\
Sgr A*					&	17:45:37.7				&	-29:00:58						&	20						&	Galactic Center, PDR 			\\
NGC 7023				&	21:01:33.9				&	+68:10:33						&	3						&	PDR							\\
L 183					&	15:54:08.5				&	-02:52:48						&	2.5						&	Dark Cloud					\\
IRC+10216				&	09:47:57.4				&	+13:16:44						&	-26						&	C-rich Circumstellar Envelope		\\		
M17-SW					&	18:20:25.1				&	-16:11:49						&	20						&	Star Forming Region, PDR		\\
IRAS 16293				&	16:32:22.6				&	-24:28:33						&	3						&	Cold Core						\\		
S140 A					&	22:19:12.1				&	+63:18:06						&	-7.6						&	PDR							\\	
S140 B					&	22:19:17.3				&	+68:18:08						&	-7.6						&	PDR							\\
CIT 6					&	10:16:02.3				&	+30:34:18						&	-2						&	C-rich Circumstellar Envelope		\\
CB 228					&	20:51:20.5				&	+56:15:45						&	-1.6						&	Translucent Cloud				\\
G+0.18-0.04				&	17:46:11.3					&	-28:48:22						&	72						&	Galactic Center, Molecular Cloud	\\
W51e2					&	19:23:43.9				&	+14:30:35						&	55						&	Hot Core						\\
Orion Bar					&	05:35:20.6				&	-05:25:14						&	10.4						&	PDR							\\
\hline
\label{bamtargetsources}
\end{tabular}
\end{minipage}
\end{table*}

Spectra were obtained in DSB mode.  For sources with no apparent emission in the observed DSB spectra, only a single IF setting was observed and averaged to produce the spectra.  For W51e2 and the Orion Bar, where signal was observed near the expected $l$-C$_3$H$^+$ frequency, at least 3 IF frequency settings were observed to isolate the signal in either the signal or image side band.  In the case of a further four sources - NGC 7023, IRC+10216, M17-SW, and IRAS 16293, sufficient IF settings were obtained to perform a full deconvolution of the data.  Details of the methods used for the deconvolution, as well as an example script, are given in \citet{McGuire2013b}.  In most cases, the expected linewidths were significantly broader than the resolution of the observations.  In these cases, the data were Hanning smoothed, normally to a resolution of $\sim$1.6 km s$^{-1}$.  A summary is given in Table~\ref{obssummary}.

With the exception of the Sgr B2(N) observations, detailed baseline fitting and subtraction was performed for each observation.  In some cases, extreme baseline structure was observed, necessitating the use of high-order polynomials to remove the ripple.  In these cases, the frequency windows for the $l$-C$_3$H$^+$ transitions were carefully examined prior to the subtraction to ensure that no potential signal from $l$-C$_3$H$^+$ was affected by the subtraction.  In the case of Sgr B2(N), where line confusion dominates the spectrum and little to no baseline is visible, a constant offset was corrected for by eye, resulting in absolute intensity uncertainties of $\sim$0.1 K - 0.2 K (see \citet{McGuire2013b} for further details).\footnote{The Sgr B2(N) observations presented here are part of a broader line survey of Sgr B2(N) from  260 - 286 GHz presented in \citet{McGuire2013b}.  The complete preliminary reduction is accessible at http://www.cv.nrao.edu/$\sim$aremijan/SLISE/.}  

\subsection{Unbiased Molecular Line Survey}
\label{slwwcso}

 The source positions and velocities used in the unbiased molecular line surveys are given in Table \ref{slwwtargetsources}.  System temperatures were generally $<$400 K during observations, with the maximum $T_{sys}$ during high opacity being $\sim$1100 K.

Two receivers and spectrometers were used for these observations.  First, a prototype 230 GHz wideband receiver \citep{Kaul04,Rice03} was used with the facility acousto-optical spectrometer (AOS) to give spectra with 4 GHz bandwidth and $\sim$0.65 MHz channel width.  Second, the facility 230 GHz wideband receiver \citep{Kooi07} was used with the facility FFTS to give spectra with 4 GHz bandwidth and $\sim$0.27 MHz channel width. Rest frequencies of 223.192 -- 251.192 GHz were used, with a 4 GHz separation between frequency settings.  IF offsets of 4.254, 6.754, 5.268, and 7.795 GHz were applied to each rest frequency.  Additional IF offsets of 6.283, 4.753, 5.767, and 7.269 GHz were applied to the two lowest rest frequency settings on each source to ensure a minimum frequency sampling redundancy of 6.  Most frequencies were sampled by 8 separate frequency settings to enable deconvolution of the DSB spectra.

The raw data were intensity calibrated using the standard chopper wheel calibration method, which placed the intensities on the atmosphere-corrected temperature scale, $T_a^{\ast}$.  A chopping frequency of 1.1 Hz was used with a chopper throw of either 70$\pm$8\arcsec or 90$\pm$8\arcsec.  A noise level of $\leq$30 mK was achieved by adjusting integration times based on the $T_{sys}$ value determined for each frequency setting.  Pointing offsets were checked at a minimum of every two hours and were consistent to $\leq$5\arcsec ~each night.  Each spectrum was also compared to previous spectra for intensity consistency as an independent verification of the pointing accuracy.  The 230 GHz full-width-half-power beam size was 33.4\arcsec~ for the prototype receiver, and 35.54\arcsec~ for the facility receiver.

The CLASS software package included in the GILDAS suite of programs (Institut de Radioastronomie Millim\'{e}trique, Grenoble, France) was used for the data reduction and deconvolution. A first degree baseline function was used to remove baselines from the DSB spectra.  Spurious noise features were removed by blanking the affected channels prior to deconvolution.  The cleaned and baseline subtracted spectra were resampled with a 1 MHz uniform channel spacing.  The standard CLASS deconvolution routine was used to deconvolve the spectra.  The initial deconvolution assumed no gain variations between the sidebands.  A second deconvolution was then constrained using this first result, with the sideband gains being allowed to vary.  The strong spectral features (i.e. those with intensities $>$2 K) were masked during deconvolution to prevent the introduction of spurious features.  These features were added back into the spectrum after deconvolution.  All intensities were then set to the main beam temperature scale, $T_{mb}$, where $T_{mb}=T_a^{\ast}/\eta_{mb}$; the main beam efficiency was determined through observations of planets to be $\eta_{mb}$ = 60 $\pm$ 9\%  for both receivers. The noise level in the final spectra is $\leq$25 mK on the $T_{mb}$ scale.  The deconvolved spectra in the frequency range covering the $l$-C$_3$H$^+$ lines are shown in Figures \ref{slww1}~--~\ref{slww3}. 

\begin{table*}
\centering
\begin{minipage}{110mm}
\caption{Sources, coordinates, $V_{LSR}$, and source type for each observed source from the unbiased line surveys.}
\begin{tabular}{l r r c l}
\hline
Source					&	$\alpha$(J2000)			&	$\delta$(J2000)					&	$V_{LSR}$ (km s$^{-1}$)		&	Notes						\\
\hline
%W3                 			&    02:27:04.61   				&   +61:52:25.0  					&    -47.0  						& Hot Core                 \\
L1448 MM-1         			&    03:25:38.80   				&   +30:44:05.0  					&      0.0  						& Class 0 + outflow        \\
NGC 1333 IRAS 2A   		&    03:28:55.40   				&   +31:14:35.0  					&      7.8  						& Hot Corino               \\
%NGC 1333 IRAS 2B   		&    03:28:57.24   				&   +31:11:14.0  					&      7.0  						& Hot Corino               \\
NGC 1333 IRAS 4A   		&    03:29:10.30  				&   +31:13:31.0  					&      6.8  						& Hot Corino               \\
NGC 1333 IRAS 4B   		&    03:29:11.99   				&   +31:13:08.9  					&      5.0  						& Hot Corino               \\
%B1-b             				&    03:33:20.80   				&   +31:07:40.0  					&      0.39 						& Class 0                  \\
Orion-KL          				&    05:35:14.16   				&   -05:22:21.5  					&      8.0  						& Hot Core                 \\
NGC 2264           			&    06:41:12.00   				&   +09:29:09.0  					&      7.6  						& Hot Core                 \\
NGC 6334-29         			&    17:19:57.00  				&   -35:57:51.0  					&     -5.0  						& Class 0                  \\
NGC 6334-38         			&    17:20:18.00  				&   -35:54:42.0  					&     -5.0  						& Class 0                  \\
NGC 6334-43         			&    17:20:23.00  				&   -35:54:55.0  					&     -2.6  						& Class 0                  \\
NGC 6334-I(N)       			&    17:20:55.00  				&   -35:45:40.0  					&     -2.6  						& Class 0                  \\
Sgr B2(N-LMH)      			&    17:47:19.89  				&   -28:22:19.3  					&     64    						& Hot Core                 \\
%GCM+0.693-0.027    		&    17:47:21.86  				&   -28:21:27.1  					&     68.0  						& Shocked region           \\ 
GAL 10.47+0.03    			&    18:08:38.40  				&   -19:51:51.8  					&     67.8  						& HII region               \\
GAL 12.21-0.10     			&    18:12:39.70  				&   -18:24:20.9  					&     24.0  						& HII region               \\                           
GAL 12.91-00.26    			&    18:14:39.00  				&   -17:52:0.30  					&     37.5  						& Hot Core                 \\
HH 80-81           			&    18:19:12.30  				&   -20:47:27.5  					&     12.2  						& Outflow                  \\
GAL 19.61-0.23     			&    18:27:37.99  				&   -11:56:42.0  					&     40.0  						& Hot Core                 \\ 
GAL 24.33+0.11 MM1   		&    18:35:08.14  				&   -07:35:01.1  					&    113.4  					& Hot Core                 \\
GAL 24.78+0.08     			&    18:36:12.60  				&   -07:11:11.0  						&    111.0  					& Hot Core                 \\
%GAL 29.96-0.02     			&    18:46:03.92  				&   -02:39:21.9  					&     97.4  						& Hot Core                 \\
GAL 31.41+0.31     			&    18:47:34.61  				&   -01:12:42.8  					&     97.0  						& Hot Core                 \\
GAL 34.3+0.20     			&    18:53:18.54  				&   +01:14:57.9  					&     58.0  						& Hot Core                 \\
GAL 45.47+0.05     			&    19:14:25.60  				&   +11:09:26.0  					&     62.0  						& Hot Core                 \\
%W51                			&    19:23:43.77  				&   +14:30:25.9  					&     55.0  						& Hot Core                 \\
GAL 75.78+0.34     			&    20:21:44.09  				&   +37:26:39.8  					&      4.0  						& HII region               \\
W75N               				&    20:38:36.60 				&   +42:37:32.0  					&     10.0  						& Hot Core                 \\
DR21(OH)           			&    20:39:01.10  				&   +42:22:49.1  					&     -3.0  						& Hot Core                 \\
L1157-MM           			&    20:39:06.20  				&   +68:02:16.0  					&      2.7  						& Class 0 + outflow        \\     
%NGC 7538           			&    23:13:45.70  				&   +61:28:21.0  					&    -57.0  						& Hot Core                 \\  
\hline
\label{slwwtargetsources}
\end{tabular}
\end{minipage}
\end{table*}

\section{Results and Data Analysis}
\label{results}

\begin{figure}
\centering
\includegraphics[width=0.45\textwidth]{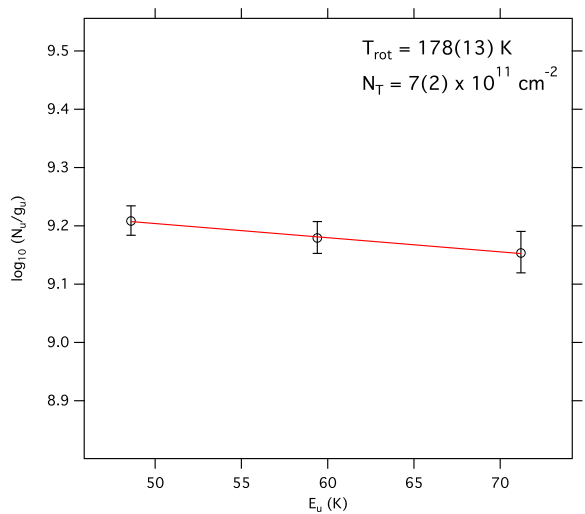}
\caption{Rotation diagram of $l$-C$_3$H$^+$ transitions observed in the Orion Bar PDR.}
\label{rots}
\end{figure}

\begin{figure*}
\centering
\begin{minipage}{175mm}
\includegraphics[width=175mm]{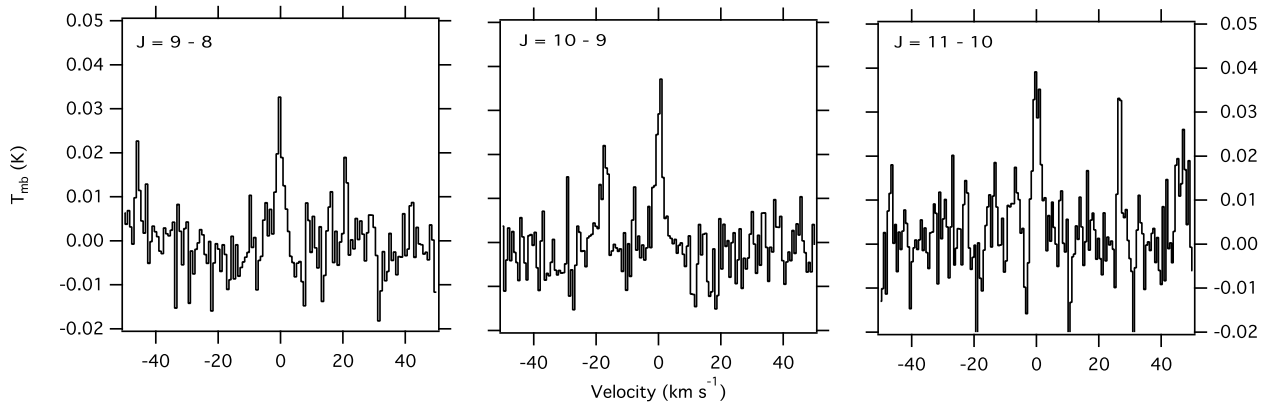}
\caption{$J = 9 -8$, $10-9$, and $11-10$ transitions of $l$-C$_3$H$^+$ observed toward the Orion Bar PDR.  The spectra are corrected for an observed source LSR velocity of 10.4 km s$^{-1}$ and have been baseline subtracted and Hanning smoothed to a resolution of 488 kHz ($\sim$0.7 km s$^{-1}$).}
\label{orionbar}
\end{minipage}
\end{figure*}

Of the sources searched here, signal from $l$-C$_3$H$^+$ was observed only toward the Orion Bar PDR (see Figure \ref{orionbar}).  Gaussian fits to the emission lines show an average FWHM width of 3.6 km s$^{-1}$ with peak intensities of 28.5 mK ($J = 9-8$), 32.9 mK ($J = 10-9$), and 37.5 mK ($J = 11-10$) and signal-noise-ratios of 4.4, 5.2, and 5.0, respectively, with a $V_{LSR} = 10.4$ km s$^{-1}$.  These values for linewidth and velocity are consistent with other molecules associated with the Orion Bar PDR \citep{Fuente2003}.  A rotational diagram analysis indicates a rotational temperate of 178(13) K and a column density of $7(2) \times 10^{11}$ cm$^{-2}$ (see Figure \ref{rots}).  A detailed examination of the rotational diagram method, as well as the equations (detailed below) used to determine column densities, can be found in \citep{Goldsmith1999}.

\begin{figure}
\centering
\includegraphics[width=80mm]{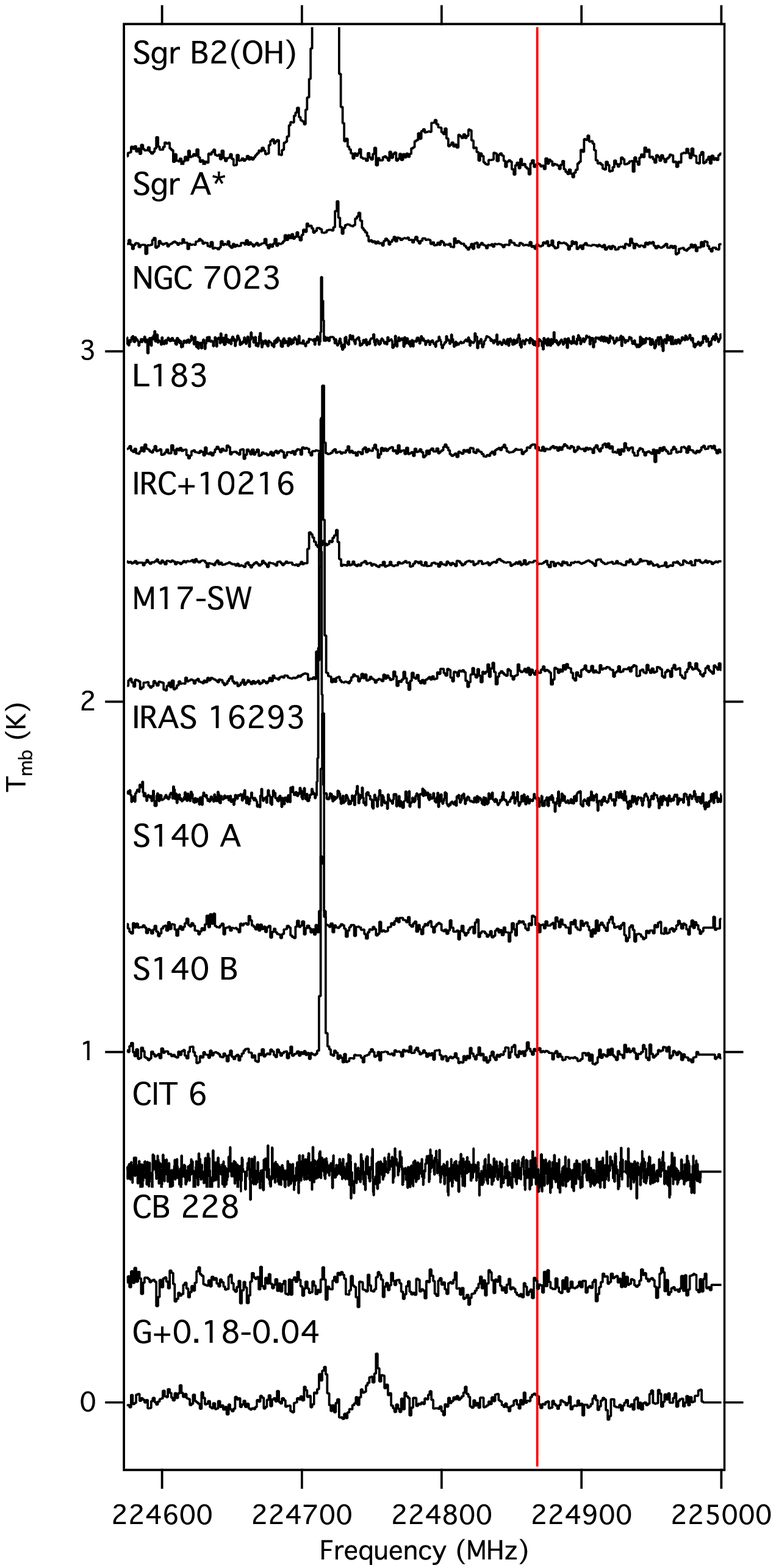}
\caption{$J = 10 -9$ spectral window toward target sources.  All spectra are adjusted to the $V_{LSR}$ indicated in Table \ref{bamtargetsources} and are vertically offset for clarity.  The feature at 224714 MHz is due to C$^{17}$O. The red vertical line indicates the frequency of the $J = 10 - 9$ transition.}
\label{10-9}
\end{figure}

\begin{figure}
\centering
\includegraphics[width=80mm]{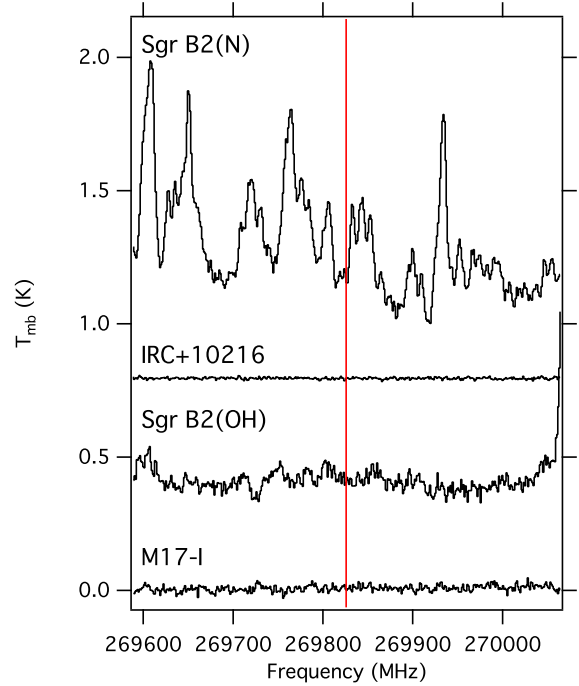}
\caption{$J = 12-11$ spectral window toward target sources.  All spectra are adjusted to the $V_{LSR}$ indicated in Table \ref{bamtargetsources} and are vertically offset for clarity. The red vertical line indicates the frequency of the $J = 12 - 11$ transition.}
\label{12-11}
\end{figure}

\begin{figure}
\centering
\includegraphics[width=80mm]{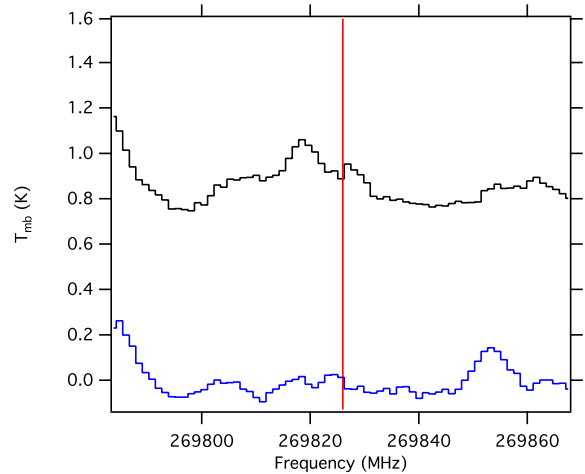}
\caption{$J = 12-11$ spectral window toward W51e2 in two different IF settings.  Spectra are DSB, adjusted to a $V_{LSR} =+55$ km s$^{-1}$, and are vertically offset for clarity.  The red vertical line indicates the frequency of the $J = 12 - 11$ transition.}
\label{w51}
\end{figure}

\begin{figure}
\centering
\includegraphics[width=80mm]{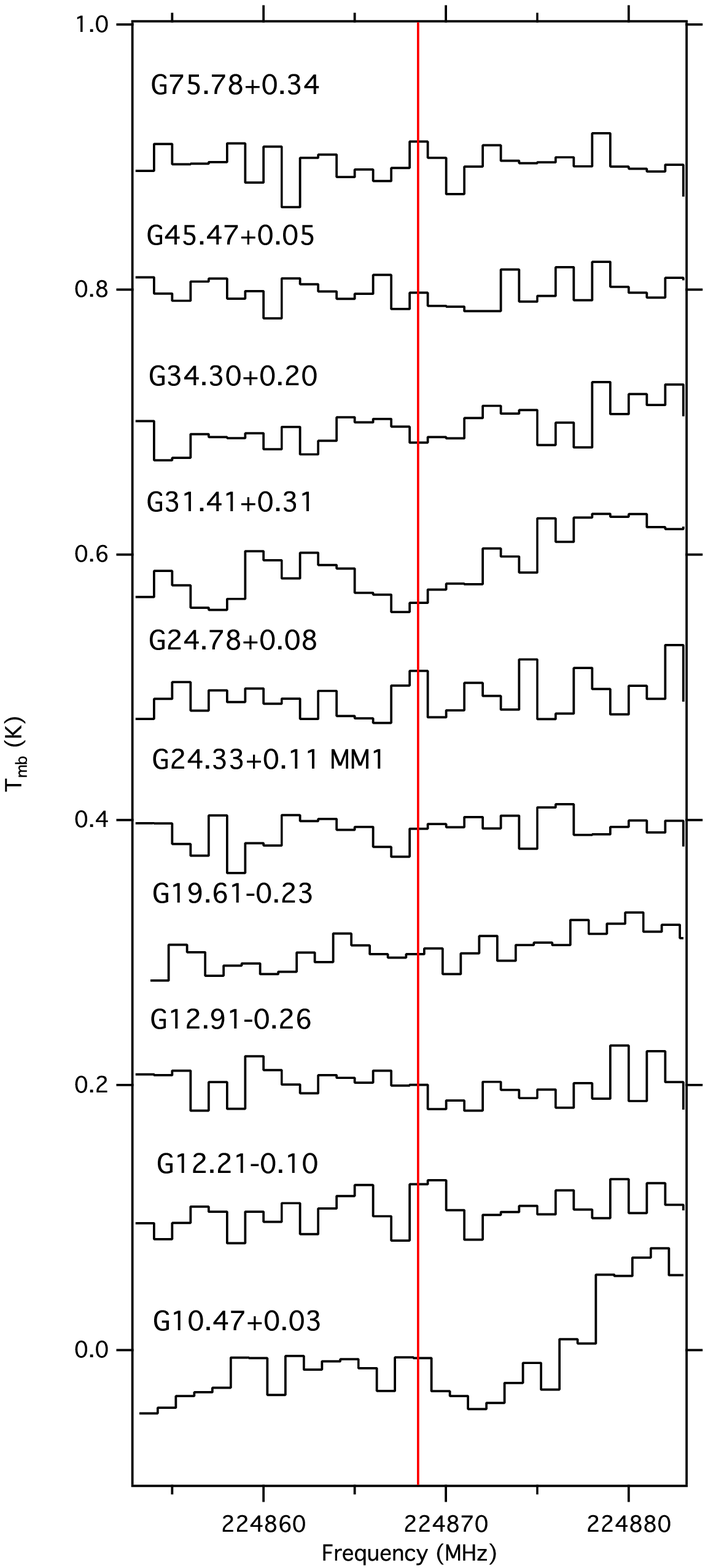}
\caption{$J = 10 -9$ spectral window toward unbiased line survey sources.  All spectra are adjusted to the $V_{LSR}$ indicated in Table \ref{slwwtargetsources} and are vertically offset for clarity.  The red vertical line indicates the frequency of the $J = 10 - 9$ transition.}
\label{slww1}
\end{figure}

\begin{figure}
\centering
\includegraphics[width=80mm]{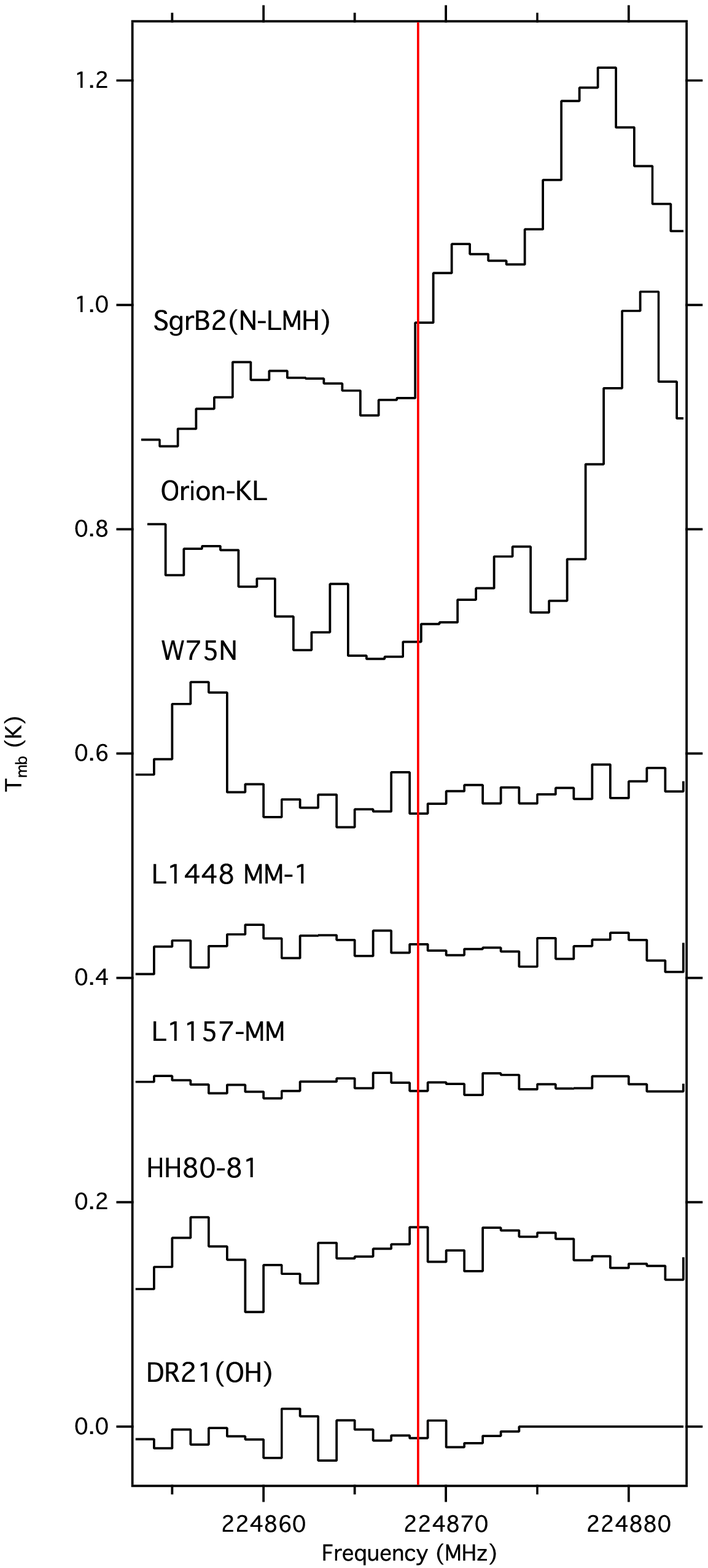}
\caption{$J = 10 -9$ spectral window toward unbiased line survey sources.  All spectra are adjusted to the $V_{LSR}$ indicated in Table \ref{slwwtargetsources} and are vertically offset for clarity.  The red vertical line indicates the frequency of the $J = 10 - 9$ transition.}
\label{slww2}
\end{figure}

\begin{figure}
\centering
\includegraphics[width=80mm]{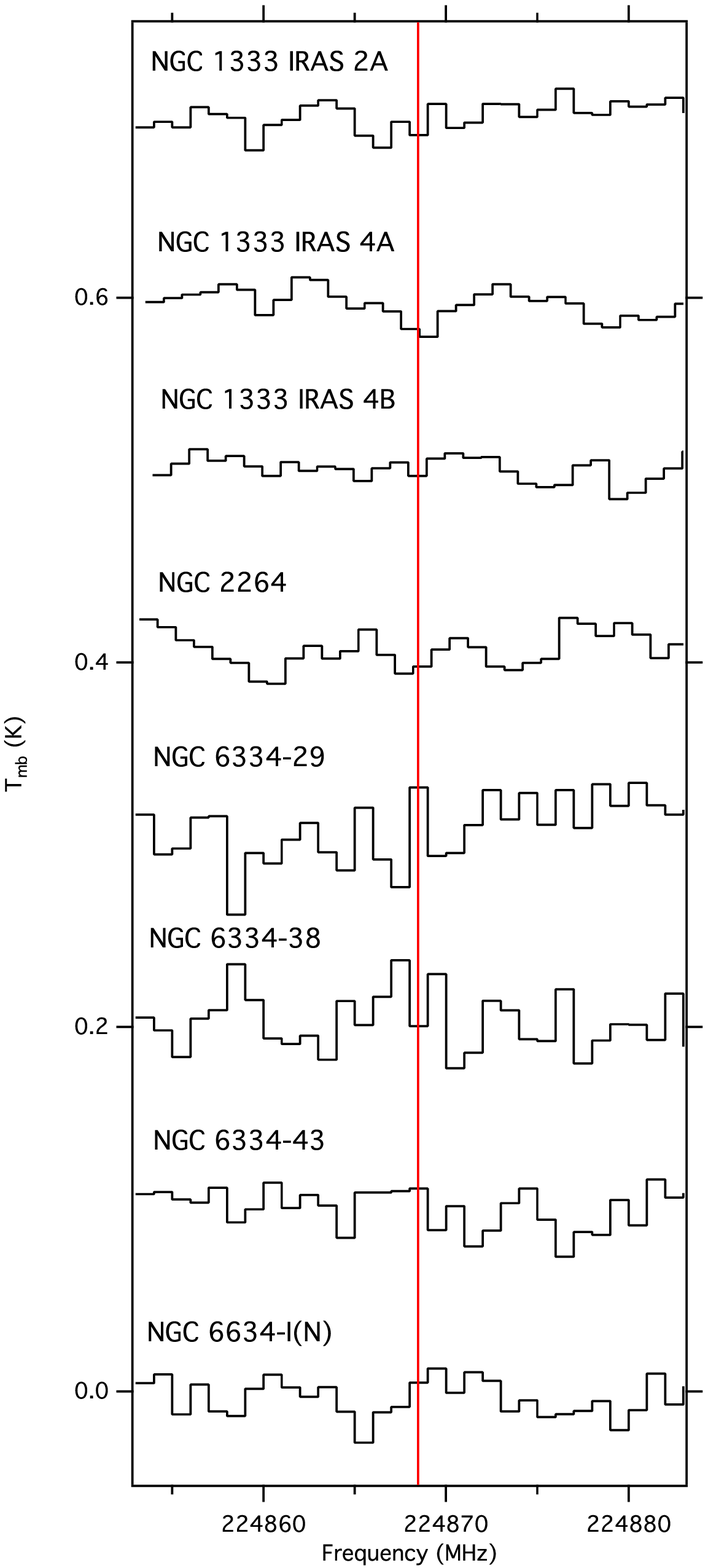}
\caption{$J = 10 -9$ spectral window toward unbiased line survey sources.  All spectra are adjusted to the $V_{LSR}$ indicated in Table \ref{slwwtargetsources} and are vertically offset for clarity.  The red vertical line indicates the frequency of the $J = 10 - 9$ transition.}
\label{slww3}
\end{figure}

The spectra collected in the targeted search, other than in the Orion Bar PDR, are shown in Figure \ref{10-9} ($J = 10 -9$) and Figures \ref{12-11} and \ref{w51} ($J=12-11$).  Spectra from the unbiased line surveys around the $J=10-9$ transition are shown in Figures \ref{slww1}~-~\ref{slww3}. 

Upper limits to the column density in each source are calculated using Equation \ref{emissioncd}, following the convention of \citet{Hollis2004}.

\begin{equation}
\label{emissioncd}
N_T=\frac{3k}{8\pi^3}\times\frac{Q_re^{E_u/T_{ex}}}{\nu S\mu ^2}\times\frac{\sqrt{\pi}}{2ln2}\times\frac{\Delta T_{mb} \Delta V/\eta _b}{1-\frac{(e^{h\nu /kT_{ex}}-1)}{(e^{h\nu /kT_{bg}}-1)}}\mbox{ cm}^{-2}
\end{equation}

Here, $N_T$ is the total column density, $Q_r$ is the rotational partition function, $E_u$ is the upper state energy, $T_{ex}$ is the excitation temperature, $\nu$ is the frequency of the transition, $S\mu^2$ is the transition strength (with $\mu$ taken as 3 Debye for $l$-C$_3$H$^+$ \citep{Pety2012}), $\Delta T_{mb}$ is the peak line intensity, $\Delta V$ is the line width, $\eta _b$ is the beam efficiency at frequency $\nu$, and $T_{bg}$ is the background temperature.  The source of the emission is assumed to completely fill the $\sim$30$\arcsec$ beam.  

For all sources in the targeted search, $\Delta T_{mb}$ was taken as the RMS noise of the appropriately smoothed spectrum and $\Delta V$ was typically determined by a Gaussian fit to the nearby C$^{17}$O line.  In some cases, such as the clearly masing Sgr A* signal or completely empty spectra, a literature value was used (see notes in Table \ref{obssummary}). Partition functions were calculated using Equation \ref{linearq} (cf. \citealt{Gordy1984}) as described by McGuire et al. (2014).

\begin{equation}
Q_r (l\mbox{-C}_3\mbox{H}^+) \approx \frac{kT}{hB} = 1.85(T)
\label{linearq}
\end{equation}

To calculate upper limits, we use the molecule-specific parameters given in \citet{Pety2012} and the upper limit $\Delta T_{mb}$ and $\Delta V$ values given in Table \ref{obssummary}.  For these frequency ranges at the CSO, $\eta _b$ is $\sim$0.70.   Upper limits for each molecule in these sources, near the two extremes of temperature so far attributed to $l$-C$_3$H$^+$, are shown in Table \ref{upperlimits}.

\begin{table*}
\centering
\begin{minipage}{140mm}
\caption{Summary of observations of the  $J = 10 -9$ and $J = 12 -11$ frequency windows.}
\begin{tabular}{l c c c c c c c c c}
\hline
					& \multicolumn{3}{c}{$J = 10 - 9$}			 & \multicolumn{3}{c}{$J = 12 - 11$}				&			&						&			\\
					& RMS		& \multicolumn{2}{c}{Resolution}& RMS 		& \multicolumn{2}{c}{Resolution}	& $\Delta V$	&						&			\\
\cline{2-4}\cline{5-7}
Source 				& (mK) 		& (MHz)		& (km s$^{-1}$)	&  (mK)	 	& (MHz) 		& (km s$^{-1}$) 	& (km s$^{-1}$)	& Switching 				& DSB/SSB 	\\
\hline
Sgr B2(OH)			&	27.5		&	1.2		&	1.6		&	32.0		&	1.2		&	1.3			& 25			&	PS					&	DSB		\\
Sgr A*				&	6.0		&	1.2		&	1.6		&	...		&	...		&	...			& 20$^a$		&	PS					&	DSB		\\
NGC 7023			&	8.3		&	0.6		&	0.8		&	...		&	...		&	...			& 2			&	Chop				&	SSB		\\
L 183				&	8.7		&	1.2		&	1.6		&	...		&	...		&	...			& 3			&	PS					&	DSB		\\
IRC+10216			&	5.8		&	1.2		&	1.6		&	3.7		&	1.2		&	1.3			& 30			&	Chop$^{\dagger}$		&	SSB		\\
M17-SW				&	10.9		&	1.2		&	1.6		&	13.0		&	1.3		&	1.4			& 5			&	PS$^{\ddagger}$		&	SSB		\\
IRAS 16293			&	11.4		&	0.6		&	0.8		&	...		&	...		&	...			& 4			&	Chop				&	SSB		\\
S140 A				&	14.1		&	1.2		&	1.6		&	...		&	...		&	...			& 4			&	Chop$^{\dagger}$		&	DSB		\\
S140 B				&	11.5		&	1.2		&	1.6		&	...		&	...		&	...			& 5			&	PS					&	DSB		\\
CIT 6				&	15.4		&	0.3		&	0.4		&	...		&	...		&	...			& 30$^b$		&	PS					&	DSB		\\
CB 228				&	21.9		&	1.2		&	1.6		&	...		&	...		&	...			& 1$^c$		&	PS					&	DSB		\\
G+0.18-0.04			&	15.4		&	1.2		&	1.6		&	...		&	...		&	...			& 27			&	PS					&	DSB		\\
W51e2				&	...		&	...		&	...		&	46.2		&	1.2		&	1.3			& 13			&	PS					&	DSB$^{\star}$  \\
Sgr B2(N)				&	...		&	...		&	...		&	15.0		&	1.2		&	1.3			& 13			&	PS					&	SSB		\\	
					&			&			&			&			&			&				& 			&						&			\\
%W3                 		&			&	1.0		&	1.3		&			&			&				& 			&	Chop				&	SSB		\\	
L1448 MM-1         		&	10.9		&	1.0		&	1.3		&	...		&	...		&	...			& 1.4			&	Chop				&	SSB		\\
NGC 1333 IRAS 2A   	&	7.7		&	1.0		&	1.3		&	...		&	...		&	...			& 3.8			&	Chop				&	SSB		\\	
%NGC 1333 IRAS 2B   	&	10.8		&	1.0		&	1.3		&			&			&				& 			&	Chop				&	SSB		\\	
NGC 1333 IRAS 4A   	&	8.5		&	1.0		&	1.3		&	...		&	...		&	...			& 5			&	Chop				&	SSB		\\	
NGC 1333 IRAS 4B   	&	9.5		&	1.0		&	1.3		&	...		&	...		&	...			& 4.1			&	Chop				&	SSB		\\	
%B1-b             			&			&	1.0		&	1.3		&			&			&				& 			&	Chop				&	SSB		\\
Orion-KL          			&	36.9		&	1.0		&	1.3		&	...		&	...		&	...			& 6.5			&	Chop				&	SSB		\\	
NGC 2264          		&	12		&	1.0		&	1.3		&	...		&	...		&	...			& 3.8			&	Chop				&	SSB		\\ 	
NGC 6334-29        		&	19.4		&	1.0		&	1.3		&	...		&	...		&	...			& 4.5			&	Chop				&	SSB		\\ 	
NGC 6334-38        		&	15.5		&	1.0		&	1.3		&	...		&	...		&	...			& 3.4			&	Chop				&	SSB		\\ 	
NGC 6334-43        		&	10.9		&	1.0		&	1.3		&	...		&	...		&	...			& 3.2			&	Chop				&	SSB		\\	
NGC 6334-I(N)       		&	10.6		&	1.0		&	1.3		&	...		&	...		&	...			& 4.8			&	Chop				&	SSB		\\	
Sgr B2(N-LMH)      		&	32.3		&	1.0		&	1.3		&	...		&	...		&	...			& 18			&	Chop				&	SSB		\\	
%GCM+0.693-0.027    	&			&	1.0		&	1.3		&			&			&				& 			&	Chop				&	SSB		\\	
GAL 10.47+0.03    		&	34.4		&	1.0		&	1.3		&	...		&	...		&	...			& 8.7			&	Chop				&	SSB		\\
GAL 12.21-0.10     		&	14.2		&	1.0		&	1.3		&	...		&	...		&	...			& 7.4			&	Chop				&	SSB		\\
GAL 12.91-00.26    		&	12.9		&	1.0		&	1.3		&	...		&	...		&	...			& 4.2			&	Chop				&	SSB		\\
HH 80-81           		&	38.2		&	1.0		&	1.3		&	...		&	...		&	...			& 2.6			&	Chop				&	SSB		\\
GAL 19.61-0.23     		&	13.6		&	1.0		&	1.3		&	...		&	...		&	...			& 7.4			&	Chop				&	SSB		\\
GAL 24.33+0.11 MM1   	&	14.2		&	1.0		&	1.3		&	...		&	...		&	...			& 4			&	Chop				&	SSB		\\	
GAL 24.78+0.08     		&	16.3		&	1.0		&	1.3		&	...		&	...		&	...			& 5.2			&	Chop				&	SSB		\\
%GAL 29.96-0.02     		&			&	1.0		&	1.3		&			&			&				& 			&	Chop				&	SSB		\\
GAL 31.41+0.31     		&	24.8		&	1.0		&	1.3		&	...		&	...		&	...			& 6.3			&	Chop				&	SSB		\\
GAL 34.3+0.20    		&	15.1		&	1.0		&	1.3		&	...		&	...		&	...			& 6.5			&	Chop				&	SSB		\\
GAL 45.47+0.05     		&	10.5		&	1.0		&	1.3		&	...		&	...		&	...			& 4.8			&	Chop				&	SSB		\\
%W51                		&			&	1.0		&	1.3		&			&			&				& 			&	Chop				&	SSB		\\
GAL 75.78+0.34     		&	13.4		&	1.0		&	1.3		&	...		&	...		&	...			& 3.5			&	Chop				&	SSB		\\
W75N               			&	17.4		&	1.0		&	1.3		&	...		&	...		&	...			& 4			&	Chop				&	SSB		\\
DR21(OH)           		&	13.5		&	1.0		&	1.3		&	...		&	...		&	...			& 6.5			&	Chop				&	SSB		\\
L1157-MM           		&	6.2		&	1.0		&	1.3		&	...		&	...		&	...			& 5.5			&	Chop				&	SSB		\\
%NGC 7538			&			&	1.0		&	1.3		&			&			&				& 			&	Chop				&	SSB		\\	
\hline
\multicolumn{10}{l}{$^{\dagger}$ At least one observation was taken in position switched mode to determine whether extended structure was being}\\
\multicolumn{10}{l}{ chopped into with the secondary mirror.  No difference was observed between the position switched and chopped off position.}\\
\multicolumn{10}{l}{$^{\ddagger}$ The throw for this source was 5$^{\prime}$}\\
\multicolumn{10}{l}{$^{\star}$ Two IF settings were observed for this source.}\\
\multicolumn{10}{l}{References -- (a) \citet{Martin2012}; (b) \citet{Chau2012}; \citet{Morisawa2005}}\\
\label{obssummary}
\end{tabular}
\end{minipage}
\end{table*}

\begin{table}
\centering
\caption{Upper limits for $l$-C$_3$H$^+$ in each source at 15 K and at 180 K.}
\begin{tabular}{l c c}
\hline
					& \multicolumn{2}{c}{$N_T$ (10$^{12}$ cm$^{-2}$)}				\\
Source				& 	15 K				& 	180 K			\\
\hline
Sgr B2(OH)			&	13				&	4.1	\\
Sgr A*				&	2.2				&	0.7	\\
NGC 7023			&	0.3				&	0.1	\\
L 183				&	0.5				&	0.2	\\
IRC+10216			&	3.3				&	1.0	\\
M17-SW				&	1.0				&	0.3	\\
IRAS 16293			&	0.9				&	0.3	\\
S140 A				&	1.1				&	0.3	\\
S140 B				&	1.1				&	0.3	\\
CIT 6				&	8.7				&	2.7	\\
CB 228				&	0.4				&	0.1	\\
G+0.18-0.04			&	7.8				&	2.5	\\
W51e2				&	41				&	2.8	\\
Sgr B2(N)				&	13				&	0.9	\\	
					&					&		\\
L1448 MM-1         		&	0.3				&	0.1	\\
NGC 1333 IRAS 2A   	&	0.5				&	0.2	\\	
NGC 1333 IRAS 4A   	&	0.8				&	0.3	\\	
NGC 1333 IRAS 4B  	&	0.7				&	0.2	\\	
Orion-KL          			&	4.5				&	1.4	\\	
NGC 2264          		&	0.9				&	0.3	\\ 	
NGC 6334-29        		&	1.6				&	0.5	\\ 	
NGC 6334-38        		&	1.0				&	0.3	\\ 	
NGC 6334-43        		&	0.7				&	0.2	\\	
NGC 6334-I(N)       		&	1.0				&	0.3	\\	
Sgr B2(N-LMH)      		&	11				&	3.4	\\	
GAL 10.47+0.03    		&	5.6				&	1.8	\\
GAL 12.21-0.10     		&	2.0				&	0.6	\\
GAL 12.91-00.26    		&	1.0				&	0.3	\\
HH 80-81           		&	1.9				&	0.6	\\
GAL 19.61-0.23     		&	1.9				&	0.6	\\
GAL 24.33+0.11 MM1   	&	1.1				&	0.3	\\	
GAL 24.78+0.08     		&	1.6				&	0.5	\\
GAL 31.41+0.31     		&	2.9				&	0.9	\\
GAL 34.3+0.20    		&	1.8				&	0.6	\\
GAL 45.47+0.05     		&	0.9				&	0.3	\\
GAL 75.78+0.34     		&	0.9				&	0.3	\\
W75N               			&	1.3				&	0.4	\\
DR21(OH)           		&	1.6				&	0.5	\\
L1157-MM           		&	0.6				&	0.2	\\     \hline
\label{upperlimits}
\end{tabular}
\end{table}

\section{Discussion}
\label{discussion}

Of the 39 sources observed in this work, $l$-C$_3$H$^+$ has been detected in only a single one: the Orion Bar PDR.  This extends the list of environments in which $l$-C$_3$H$^+$ is known to be present to three: the Orion Bar PDR, the Horsehead PDR, and Sgr B2(N), with tenuous evidence for $l$-C$_3$H$^+$ in Sgr B2(OH) and TMC-1.   

The lack of detection of $l$-C$_3$H$^+$, in reasonably high-sensitivity observations, toward any molecularly-rich hot core source outside of Sgr B2(N) is initially puzzling.  The detection in Sgr B2(N) by \citet{McGuire2013a} may have been fortuitous - the highly sub-thermal nature of the observed absorption features may have allowed their observation despite an otherwise low abundance that would typically preclude detection.  Perhaps more puzzling is the lack of detection in the majority of the PDR sources observed here.  The answer is almost certainly one of temperature; with column densities similar to those found the Horsehead and Orion Bar PDRs, the observations presented here with the CSO are not sensitive to material cooler than $\sim$130 K.  Thus, further high-sensitivity observations at 3 mm, where the Boltzmann peak for cold $l$-C$_3$H$^+$ falls, are warranted to fully explore the range of excitation conditions so far attributed to this molecule.

Further insight into likely sources in which $l$-C$_3$H$^+$ could be found may also be gained by comparing its formation and destruction pathways to that of HOC$^+$.  As described by \citet{Pety2012}, the primary formation mechanism for $l$-C$_3$H$^+$ is through the reaction of acetylene with C$^+$.  Destruction readily occurs via reaction with molecular hydrogen (see Eqs. \ref{c3hpform} - \ref{c3hpdest}).

\begin{equation}
\mbox{C}_2\mbox{H}_2 + \mbox{C}^+ \rightarrow \mbox{C}_3\mbox{H}^+
\label{c3hpform}
\end{equation}

\begin{equation}
\mbox{C}_3\mbox{H}^+ \xrightarrow{\mbox{H}_2} \mbox{C}_3\mbox{H}_2^+ \xrightarrow{e^-} \mbox{C}_3\mbox{H}
\end{equation} 

\begin{equation}
\mbox{C}_3\mbox{H}^+ \xrightarrow{\mbox{H}_2} \mbox{C}_3\mbox{H}_3^+ \xrightarrow{e^-} \mbox{C}_3\mbox{H}_2
\label{c3hpdest}
\end{equation} 

The detections of $l$-C$_3$H$^+$ in the Horsehead and Orion Bar PDRs support these formation and destruction mechanisms.  Destruction via H$_2$ is expected to be rapid and thus dominate $l$-C$_3$H$^+$ populations under typical conditions.  Within PDR sources, however, where the ultraviolet radiation field is greatly enhanced relative to the mean interstellar value, sufficient C$^+$ may be present to compete with this destruction pathway and lead to detectable abundances of $l$-C$_3$H$^+$. 

A comparison can be made with the chemistry of HOC$^+$, which is also formed, directly and indirectly, through reactions of C$^+$ and destroyed by reaction with H$_2$ to form HCO$^+$ (see Eqs. \ref{hocform} - \ref{hocdest}, c.f. \citet{Smith2002,Fuente2003}).  Thus, observations of a high HOC$^+$ abundance (or alternatively an enhanced [HOC$^+$]/[HCO$^+$] ratio) may be indicative of chemical and physical conditions that will also favor the production of $l$-C$_3$H$^+$, and act as a probe to guide future sources.

\begin{equation}
\mbox{H}_2\mbox{O} \xrightarrow{\mbox{C}^+} \mbox{HOC}^+
\label{hocform}
\end{equation}

\begin{equation}
\mbox{OH} \xrightarrow{\mbox{C}^+} \mbox{CO}^+ \xrightarrow{\mbox{H}_2} \mbox{HOC}^+
\end{equation}

\begin{equation}
\mbox{HOC}^+ \xrightarrow{\mbox{H}_2} \mbox{HCO}^+
\label{hocdest}
\end{equation}

Indeed, such an enhanced HOC$^+$ abundance (and enhanced [HOC$^+$]/[HCO$^+$] ratio) has already been detected toward both the Orion Bar PDR \citep{Fuente2003} and the Horsehead PDR \citep{Goicoechea2009}, with these values peaking in the region of the PDR.  A similarly enhanced presence of HOC$^+$ is detected by \citet{Fuente2003} in observations of the NGC 7023 PDR, for which $l$-C$_3$H$^+$ is not detected in these CSO observations.  This suggests that perhaps $l$-C$_3$H$^+$ is simply too cold to be detected in our observations at this sensitivity, and that observations at lower frequencies may result in a detection.

Observations of HOC$^+$ in diffuse clouds by \citet{Liszt2004}, find abundance ratios of [HOC$^+$]/[HCO$^+$] comparable to that of the extreme cases of PDR enhancements in the Orion Bar and NGC 7023.  This is consistent with the tentative detections of $l$-C$_3$H$^+$ absorption in diffuse, spiral arm clouds by \citet{McGuire2013a} along the line of sight to Sgr B2(N), and suggests these sources may be intriguing targets for future observations.

Finally, detections of HOC$^+$ toward other regions with enhanced FUV flux provide other tantalizing sources for future study.  A small list of these sources includes, for example, the Monocerous R2 ultracompact HII region \citep{Ginard2012} and the external galaxies NGC 253  \citep{Martin2009} and M 82 \citep{Fuente2005}.

\section{Conclusions}
\label{conclusions}

We have examined the results of both a dedicated campaign targeting $l$-C$_3$H$^+$ in 14 sources, as well as 25 additional sources from unbiased molecular line surveys with frequency coverage coincident with $l$-C$_3$H$^+$ transitions.  We detect the presence of $l$-C$_3$H$^+$ in only a single source: the Orion Bar PDR.  These observations are only sensitive to relatively warm material, and follow-up observations at lower frequencies are necessary to obtain a complete picture of $l$-C$_3$H$^+$ in these sources.  Finally, a comparison to the chemistry of HOC$^+$ has shown that HOC$^+$ has the potential to serve as a tracer of $l$-C$_3$H$^+$.  Previous observations of HOC$^+$ can therefore be used as a guide to efficiently target new sources in additional searches for $l$-C$_3$H$^+$.

\section*{Acknowledgments}

The authors express their sincere gratitude to the staff of the CSO, in particular S. Radford, R. Chamberlin, E. Bufil, B. Force, H. Yoshida, and D. Bisel.  We thank the anonymous referee for helpful comments which have improved the quality of this manuscript.  BAM is funded by an NSF Graduate Research Fellowship.  Support for this work was provided in part by the National Science Foundation.  JLS was supported by SLWW's startup funds, provided by Emory University, and by the Emory Summer Undergraduate Research at Emory (SURE) program, which is partially supported by the Howard Hughes Medical Institute.  We thank N. Wehres, T. Cross, M. Radhuber, J. Laas, J. Kroll, and B. Hays for assistance in CSO observations and data reduction.  We thank M. Sumner, F. Rice, and J. Zmuidzinas for technical support with the prototype receiver and assistance in data collection for the Orion-KL spectrum.  We thank D. Lis, M. Emprechtinger, P. Schilke, and C. Comito for helpful discussions regarding the analysis of line surveys and deconvolution of DSB spectra, and T. Phillips for his guidance and support.  We also thank the support staff from Caltech and Emory.  The CSO is operated by the California Institute of Technology under contract from the National Science Foundation. The National Radio Astronomy Observatory is a facility of the National Science Foundation operated under cooperative agreement by Associated Universities, Inc.

\bsp

\clearpage

\label{lastpage}

\end{document}